\documentclass[runningheads,a4paper]{llncs}

\usepackage{amssymb}
\usepackage{graphicx}
\usepackage{marvosym}
\usepackage{url}
\urldef{\mailsa}\path|{dyakonova,khoperskov,khrapov}@volsu.ru|
\newcommand{\keywords}[1]{\par\addvspace\baselineskip
\noindent\keywordname\enspace\ignorespaces#1}

\begin{document}

\mainmatter  

\title{Numerical Model of Shallow Water:\\the Use of NVIDIA CUDA Graphics Processors}

\titlerunning{Numerical Model of Shallow Water: the Use of GPU}

%
%
\author{Tatyana Dyakonova\textsuperscript{(\Letter)}%
\and Alexander Khoperskov\and Sergey Khrapov}
\authorrunning{T. Dyakonova, A. Khoperskov, S. Khrapov}

\institute{Volgograd State University, Volgograd, Russia\\
\mailsa\\
}

%
%

\toctitle{Lecture Notes in Computer Science}
\tocauthor{Authors' Instructions}
\maketitle

\begin{abstract}
In the paper we discuss the main features of the software package for numerical simulations of the surface water dynamics. We consider an approximation of the shallow water equations together with the parallel technologies for NVIDIA CUDA graphics processors. The numerical hydrodynamic code is based on the combined Lagrangian-Euler method~(CSPH-TVD). We focused on the features of the parallel implementation of Tesla line of graphics processors: C2070, K20, K40, K80. By using hierarchical grid systems at different spatial scales we increase the efficiency of the computing resources usage and speed up our simulations of a various flooding problems.

\end{abstract}

\section{Introduction}
 Various hydrology problems for the real terrain surface $b(x,y)$ with taking into account important physical factors for large areas can be studied by using  modern computational technologies~\cite{pisarev-khrapov2013}. Believed that control of the hydrological regime of the floodplain landscape during the spring flood on large rivers is one of the most important problem facing the numerical simulation \cite{voronin2012}.
Solution of this problem requires a very efficient numerical method based on a parallel technology~\cite{glinskiy2014}. For example, to make a both ecological and economic management of the Volga-Akhtuba floodplain we have to solve the optimization problem of hydrograph for the specific conditions of each year~\cite{Voronin2015}. We also need to explore the results of different modes of operation of tens hydraulic structures in the floodplain and for a new facilities projects. Each that research requires hundreds of numerical experiments on the basis of a direct hydrodynamic simulation of the shallow water dynamics on the area of $2000\times20000$ square kilometers.

Our practice of using large supercomputers\footnote{ in particular, the one is at Research Computing Center of M.V. Lomonosov Moscow State University \cite{Voevodin}} for a large number of hydrodynamic simulations arises a number of problems related to necessity to do a numerous simulations during the short time period and then following transfer of a large massive of data for later processing and analysis. Both performance of calculations and post-processing of the simulation data are an important factors in the usage of such models in practice. An additional problem is the visualization of the calculations, which seems a common difficulty for a very high-performance machines \cite{Moreland2015}. However we can partly solve these problems in case of using the computing resources such as personal supercomputers based on GPU accelerators. The paper discusses the results of the software package development for the parallel hydrodynamic simulations on the nodes C2070, K20, K40, K80.

\section{Mathematical and Numerical Models}

\subsection{Basic Equations}

 Numerical simulations are based on the shallow water model (Saint-Venant equations) in the following form:
\begin{equation}\label{eq-SV-H}
 \frac{\partial H}{\partial t}+\frac{\partial HU_x}{\partial x}+\frac{\partial HU_y}{\partial y}=\sigma \; ,
\end{equation}
\begin{equation}\label{eq-SV-ux}
\frac{\partial U_x}{\partial t}+U_x\frac{\partial U_x}{\partial x}+U_y\frac{\partial U_x}{\partial y}=-g\frac{\partial \eta}{\partial x}+F_x+\frac{\sigma}{H} (V_x-U_x) \; ,
\end{equation}
\begin{equation}\label{eq-SV-uy}
\frac{\partial U_y}{\partial t}+U_x\frac{\partial U_y}{\partial x}+U_y\frac{\partial U_y}{\partial y}=-g\frac{\partial \eta}{\partial y}+F_y+\frac{\sigma}{H} (V_y-U_y) \; ,
\end{equation}
where $H$~is the water depth, $U_{x}, U_{y}$~are the horizontal components of water velocity vector $\vec{U}$, which is averaged along the vertical direction, $\sigma$~is the surface density of the water sources and drains [m/sec], $g$~is gravitational acceleration, $\eta(x,y,t)=H(x,y,t)+b(x,y)$~is the free water surface level, $V_{x}, V_{y}$~are the mean horizontal velocity vector components of water at the source or drain ($\vec{V}=V_{x}\vec{e_x}+ V_{y}\vec{e_y}$), $F_{x}$, $F_{y}$~are the horizontal components of the external and internal forces ($\vec{F}=F_x\vec{e_x}+F_y\vec{e_y}$) acting the water layer.
The total density of the forces can be written as
\begin{equation}
\vec{F}=\vec{F}^{fric}+\vec{F}^{visc}+\vec{F}^{cor}+\vec{F}^{wind} \; ,
\end{equation}
where $\vec{F}^{fric}=-\frac{\lambda}{2}\vec{U}|\vec{U}|$~is the force of bottom friction, $\lambda=\frac{2gn^{2}_M}{H^{4/3}}$~is the value of hydraulic friction, $n_{M}$~is the phenomenological Manning roughness coefficient, $\vec{F}^{visc}=\nu(\frac{\partial^2U_x}{\partial x^2}+\frac{\partial^2U_y}{\partial y^2})$~is the viscous force of internal friction between layers of flow, $\nu$~is the kinematic turbulent viscosity, $\vec{F}^{cor}=2[\vec{U}\times\vec{\Omega}]$~is the Coriolis force, $\vec{\Omega}$~is the angular velocity of Earth's rotation, $\vec{f}^{wind}=C_{a}\frac{\rho_{a}}{\rho H}(\vec{W}-\vec{U})|\vec{W}-\vec{U}|$~is the wind force acting on the water layer, parameter $C_{a}$ determines the state of the water surface, $\rho_{a}$ and $\rho$~are the densities of air and water, respectively, $\vec{W}$~is the wind velocity vector in the horizontal direction.

The model (\ref{eq-SV-H})--(\ref{eq-SV-uy}) takes into account the following factors \cite{Dta2014}: irregular, inhomogeneous terrain $b(x,y)$; flow interaction with the underlying inhomogeneous topography; Earth's rotation; interaction of water flow with wind; sources, caused by work of hydraulic structures and rainfall; filtration and evaporation; internal friction due to turbulent transport.

In work \cite{glinskiy2014} proposed the so-called ``Co--design'' approach of the computational models construction. It takes into account the architecture of the supercomputer when one is creating the program code. The increased efficiency is based on the maximization of independent calculations and taking into account the peculiarities of the numerical algorithms for solving the equations of Saint-Venant in inhomogeneous terrain $b(x,y)$.

\subsection{Grids System and Matrix of Digital Terrain Elevation}
Despite the complex, irregular topography  of large rivers (Volga, Akhtuba) riverbeds, numerous channels and small ducts, in the simulations with the unsteady ``wet-dry'' type boundaries  we used an uniform Cartesian grid  $\Delta{x}_i=\Delta{y}_j=\Delta{x}=\Delta{y}$ which let us increase the efficiency of CSPH-TVD method (see paragraph~\ref{subsec-numerical}).  Typical grid size $\ell$ is limited by the depth of the fluid $H$. Since our problem is strongly non-steady, the fluid depth in computational cells can vary from 10\,cm in the flooded areas of land up to 30\,m in riverbed of the Volga. Therefore, for the shallow water model we use a large-scale grid for the simulation of deep channel areas and a small-scale grid for the calculations of flooded land areas.

\begin{figure}[!h]
\centering
\includegraphics[height=11cm]{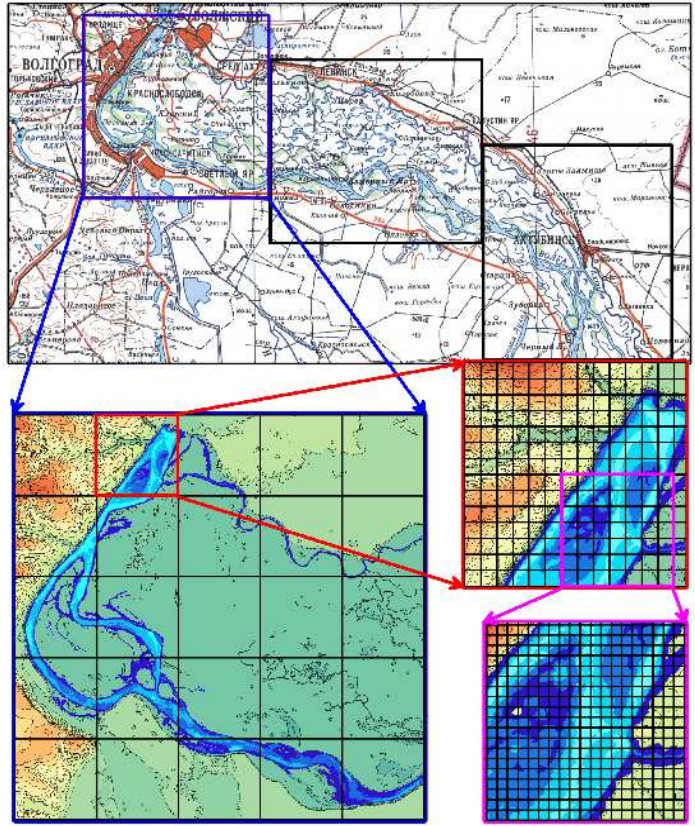}
\caption{The hierarchical system of grids for the flooded area between the Volga and Akhtuba Rivers.}
\label{Fig-grid}
\end{figure}

Figure\,\ref{Fig-grid} shows a grid structure which allows us to use efficiently computational resources because the fluid flow occurs only in a small number of cells. We simulate the dynamics of the surface water on hierarchical grid system (HGS) sequentially from the smaller to the largest scales with taking into account not smooth source distribution. Zoom-in technology is used only for mission-critical areas \cite{Bonoli2016}. This type of simulation is based on usage of two (or even more) grids with different spatial resolution depending on physical parameters of the water flow. In this case the simulated flows on the small-scale grid affects on the simulation on the larger grid. Zoom-in models without this feedback are less accurate but faster in computational sense.

\subsection{The Numerical Hydrodynamic Scheme}\label{subsec-numerical}

\begin{figure}[!h]
\centering
\includegraphics[height=7.5cm]{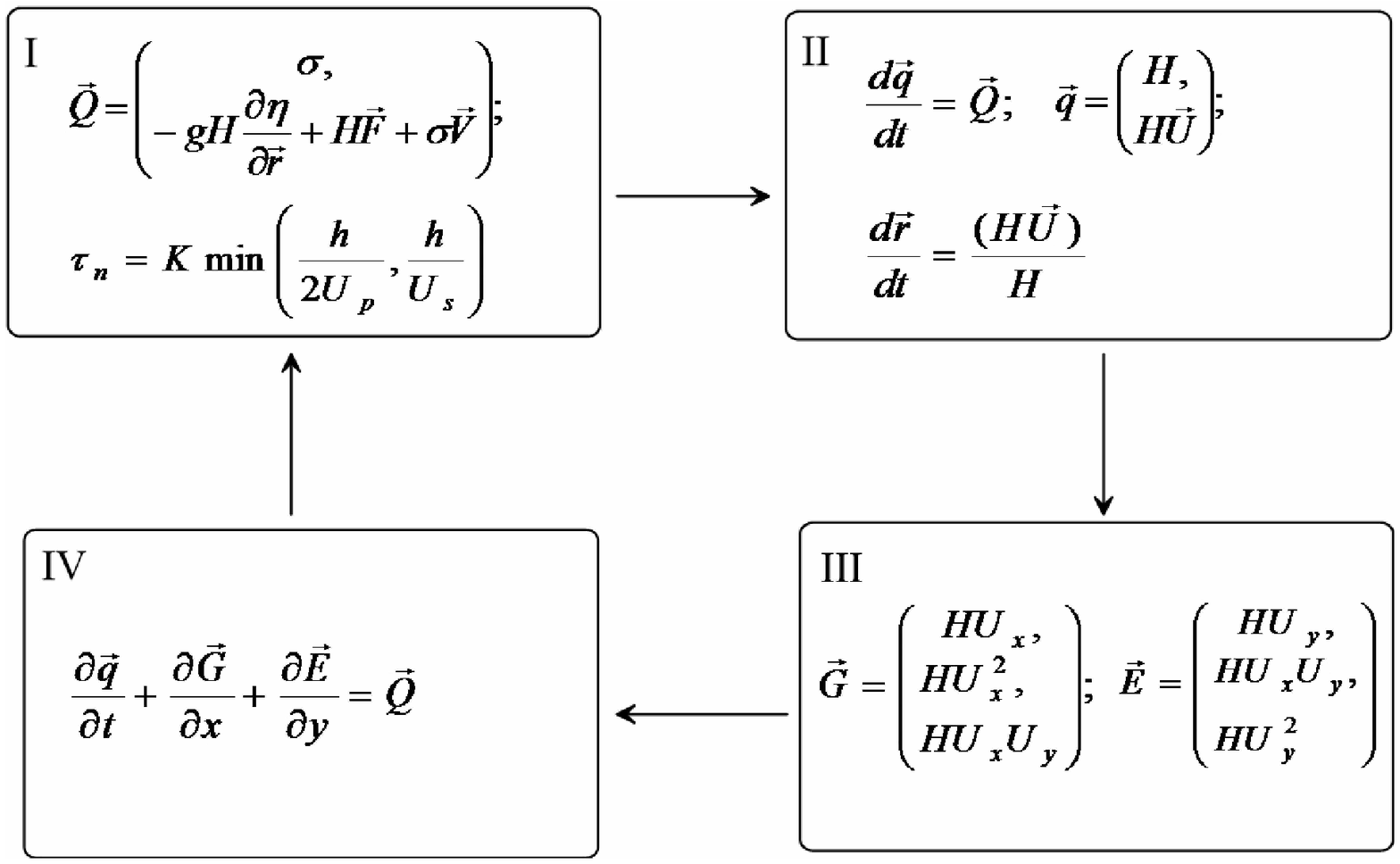}
\caption{The main stages of the computational scheme for the solution of shallow water equations.}
\label{Fig-Skhema}
\end{figure}

Figure\,\ref{Fig-Skhema} shows the calculation scheme of the CSPH-TVD method~\cite{khrapov2011,khrapov-pisarev2013,Kuzmin2014,Belousov2015}, where $h=\Delta{x}=\Delta{y}$~is the spatial resolution, $0<K<1$ is the Courant number which determines the stability of our numerical scheme, $U_p=\max[|U_x^n+\textrm{sign}(F_x)\sqrt{hF_x}|,$ $|U_y^n+\textrm{sign}(F_y)\sqrt{hF_y}|]$, $U_s=\max(|U_x^n|+\sqrt{gH^n},|U_y^n|+\sqrt{gH^n}$.
 There are four main stages of computations at a given time step $t_n$. Lagrangian approach is applied to the I and II stages, and the third and fourth stages are based on Euler approach. Both source terms $\vec{Q}$ and $\sigma$ (see Fig.\,\ref{Fig-Skhema} and (\ref{eq-SV-H})) are determined by external and internal forces respectively, and they have to be calculated firstly. Time step $\tau_n$ is also calculated at this stage.
  Then, at the second stage, we calculate the changes of variables $\vec{q}$ by using the results we obtained at the first stage. Thus, we find the displacement of the Lagrangian particles $\Delta{\vec{r}}$ inside the cells. At this stage the predictor-corrector scheme gives the second order accuracy for time integration. In the third stage, the fluxes of mass and momentum through the boundaries of Euler cells are calculated by using the approximate solution of the Riemann problem. In the last stage we update the values of $\vec{q}$ on the next time step $t_{n+1}$ and here we also put back the particles to the centers of the grid cells $(x_i, y_j)$.

\begin{figure}[!h]
\centering
\includegraphics[height=4.0cm]{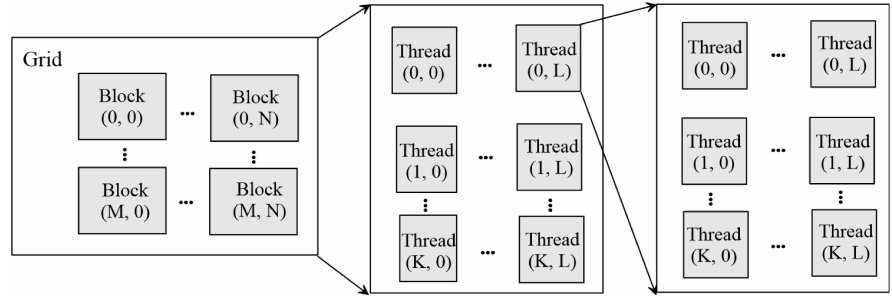}
\caption{The hierarchy of threads on the GPU devices adjusted for the dynamic parallelization.}
\label{Fig-CUDA-grid}
\end{figure}

There are several advantages of the numerical scheme described above, such as the second order of accuracy for the both time and spatial integration, the conservativeness, the well-balanced property and the straightforward simulation of a ``water -- dry bottom'' dynamical boundaries without any regularization \cite{Dta2016}.

\section{Parallel Realization of Numerical Model}

CUDA technology was used to parallelize the CSPH-TVD  numerical scheme which in turn let us to use efficiently the hierarchical grid system~(HGS, see paragraph\,2.1). This is due to the fact that HGS blocks are a kind of analogue of CUDA stream blocks, which provide the execution of CUDA-cores (Fig.\,\ref{Fig-CUDA-grid}). Using CUDA dynamic parallelism is a feature of implemented approach that allows detailed calculations of hydrodynamic flows in the small-scale grids with additional threads for the most important spatial zones associated with irregular topography.

\begin{figure}[!h]
\centering
\includegraphics[height=16.0cm]{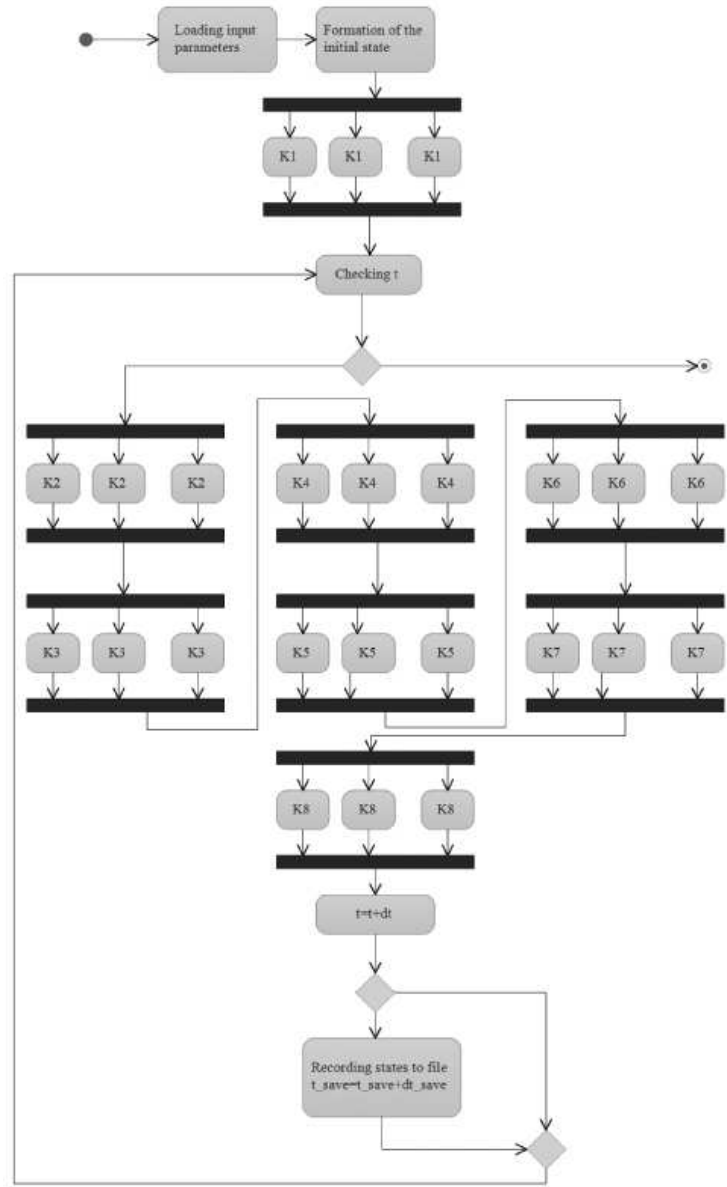}
\caption{Activity diagram for the calculation module.}
\label{Fig-Diagramma}
\end{figure}

 The computational algorithm described in paragraph 2 is parallelized by using the hybrid OpenMP-CUDA parallel programming model.
 Activity diagram of the main stages of the numerical algorithm is shown in Figure\,\ref{Fig-Diagramma}.
 We use the following notations for computing CUDA cores:
 K1 --- kernel\_Index\_block is the determination of water-filled blocks;
 K2 --- kernel\_forces\_predictor calculates the forces at the time $ t_n $ on the Lagrangian stage;
 K3 --- kernel\_dt calculates the time step $\Delta{t_{n+1}}$, depending on the flow parameters on the layer $n$;
 K4 --- kernel\_SPH\_predictor calculates the new provisions of the particles and the integral characteristics at time $t_{n+1/2}$;
 K5 --- kernel\_forces\_corrector determines the forces on the intermediate time layer $t_{n+1/2}$;
 K6 --- kernel\_SPH\_corrector calculates the positions of the particles and the integrated characteristics for the next time layer $t_{n+1}$;
 K7 --- kernel\_TVD\_flux  calculates the flux physical quantities through the cell boundaries at time $t_{n+1}$;
 K8 --- kernel\_Final determines the final hydrodynamic parameters at time $t_{n+1}$.

 The diagram in the figure\,\ref{Fig-Diagramma} demonstrates the features of CSPH--TVD method. We emphasize that the computational algorithm separation is optimal usage of GPU resources of eight CUDA-cores in case of shallow water flows on the irregular topography.

\begin{figure}[!h]
\centering
\includegraphics[height=7.9cm]{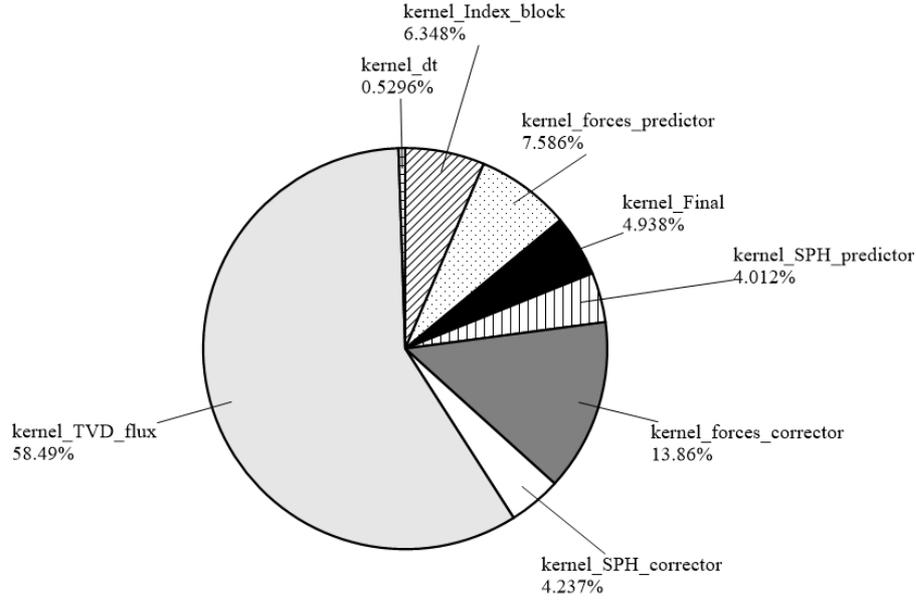}
\caption{Contributions of different stages of the CSPH-TVD numerical scheme at a given time step.}
\label{Fig-diagram}
\end{figure}

 Figure\,\ref{Fig-diagram} shows the execution time proportions of the main stages of the numerical algorithm for the corresponding CUDA-cores. We spend almost 60\,\% of the time on the calculation of the fluxes for the TVD-stage kernel\_TVD\_flux (see Fig.\,\ref{Fig-diagram}).
 This is because at this stage we solve the Riemann problem.
 Taking into account the calculation of forces kernel\_forces\_predictor and kernel\_forces\_corrector the total contribution from TVD-stage increases and it is dominant.
 We have just over 8\,\%
 from the Lagrangian stage (sum of kernel\_SPH\_predictor and kernel\_SPH\_corrector).
 Thus, the SPH-stage significantly improves the properties of the numerical scheme, but an additional stage has little effect on the final calculation time.

Fragment of the program (See below) implements the final stage of the calculation (CUDA-core K8) with checking of the presence of liquid in the CUDA-block with size of $16\times 16$ cells.
 If liquid is absent in the started block (parameter Index\_block is zero), the computations are skipped for all threads in the block. Similar approach we apply for other CUDA-cores as well (K2, K4, K5, K6 and K7).

\medskip

\noindent
{\it The code fragment for CUDA-cores K8}
\begin{verbatim}
__global__ void kernel_Final(double *H, double2 *HV, double *Ht,
       double2 *HVt, double *Fh, double2 *Fv, int2 *Index_block,
       double tau){
    int ib=blockIdx.x+blockIdx.y*gridDim.x;
    int x = threadIdx.x + blockIdx.x * blockDim.x, y = threadIdx.y +
            blockIdx.y * blockDim.y;
    int ind = x + y * blockDim.x * gridDim.x;
 if(Index_block[ib].x > 0 || Index_block[ib].y > 0){
    double dt_h = tau/dd.hp, Eps=dd.Eps, ht, Flux_h;
    double2 hv=make_double2(0,0);
    ht = Ht[ind]; Flux_h=Fh[ind];
    if( ht>Eps || fabs(Flux_h)>Eps){
        ht = dev_h(ht + dt_h*Flux_h); 	
        hv.x = dev_hv(HVt[ind].x + dt_h*Fv[ind].x,ht);	
        hv.y = dev_hv(HVt[ind].y + dt_h*Fv[ind].y,ht);
    }	
    H[ind]=ht; HV[ind]=hv; Ht[ind]=0; HVt[ind]=make_double2(0,0);
 }else {Ht[ind]=0; HVt[ind] = make_double2(0,0);}
}
\end{verbatim}

  Parameter Index\_block is determined in the CUDA-core K1 (See below).
 The variable Index\_block is the structure of type int2 containing two integer fields Index\_block.x and Index\_block.y.
  The condition Index\_block.x\,$>0$ indicates the presence of water at least in one cell of this CUDA-block.
 If the condition Index\_block.y\,$>0$ is satisfied, there is a water at least in one of the CUDA-block surrounding cells.
Since there are liquid fluxes through the CUDA-block boundaries on the Euler stage in the CUDA-core K7, we have to check the availability of water in the boundary cells of the surrounding CUDA-blocks.

\medskip

\noindent
{\it The code fragment for CUDA-core K1}
\begin{verbatim}
__global__ void kernel_Index_block(int2 *Index_block, int *Index_Q,
  double *H){
    __shared__ int2 Sij[ithbx*ithby];
    int ind_thb = threadIdx.x + ithbx*threadIdx.y;
    int ib=blockIdx.x+blockIdx.y*gridDim.x;
    int x = threadIdx.x + blockIdx.x*blockDim.x, y = threadIdx.y +
            blockIdx.y*blockDim.y;
    int tid, xx, yy, i, j, si, sj, Ni, Nj, isi, jsj, m;
    double2 D=make_double2(0,0); int2 iD=make_int2(0,0);
    if(threadIdx.x == 0){Ni=1; si=-1;}
    else if(threadIdx.x == ithbx-1){Ni=1; si=1;}
    else {Ni=0; si=0;}
    if(threadIdx.y == 0){Nj=1; sj=-1;}
    else if(threadIdx.y == ithby-1){Nj=1; sj=1;}
    else {Nj=0; sj=0;}
    for(i=0; i<=Ni; i++){
      isi = i*si;	
      if(x==0 && isi<=-1) xx = x;			
      else if(x==dd.Nx-1 && isi>=1) xx = x;			
      else xx = x + isi;
      if(i==0) m=Nj; else m=0;
      for(j=0; j<=m; j++){
        jsj = j*sj;	
        if(y==0 && jsj<=-1) yy = y;			
        else if(y==dd.Ny-1 && jsj>=1) yy = y;			
        else yy = y + jsj;
        tid = xx + yy*dd.Nx;
        if(i>0 || j>0)  {D.y += H[tid]; iD.y += Index_Q[tid]; }
        if(i==0 & j==0) {D.x += H[tid]; iD.x += Index_Q[tid]; }
      }
    }
    if(D.x>dd.Eps || iD.x>0) Sij[ind_thb].x=1;
    else Sij[ind_thb].x=0;
    if(D.y>dd.Eps || iD.y>0) Sij[ind_thb].y=1;
    else Sij[ind_thb].y=0;
    __syncthreads();
    int k = ithbx*ithby/2;
    while(k != 0){
      if(ind_thb < k) {Sij[ind_thb].x += Sij[ind_thb+k].x;
                       Sij[ind_thb].y += Sij[ind_thb+k].y; }
      __syncthreads();
      k /= 2; }
    if(ind_thb == 0) {Index_block[ib].x = Sij[0].x;
                      Index_block[ib].y = Sij[0].y; }
}
\end{verbatim}

\begin{figure}[!h]
 \vskip 0.\hsize \hskip 0.0\hsize
            \includegraphics[width=0.55\hsize]{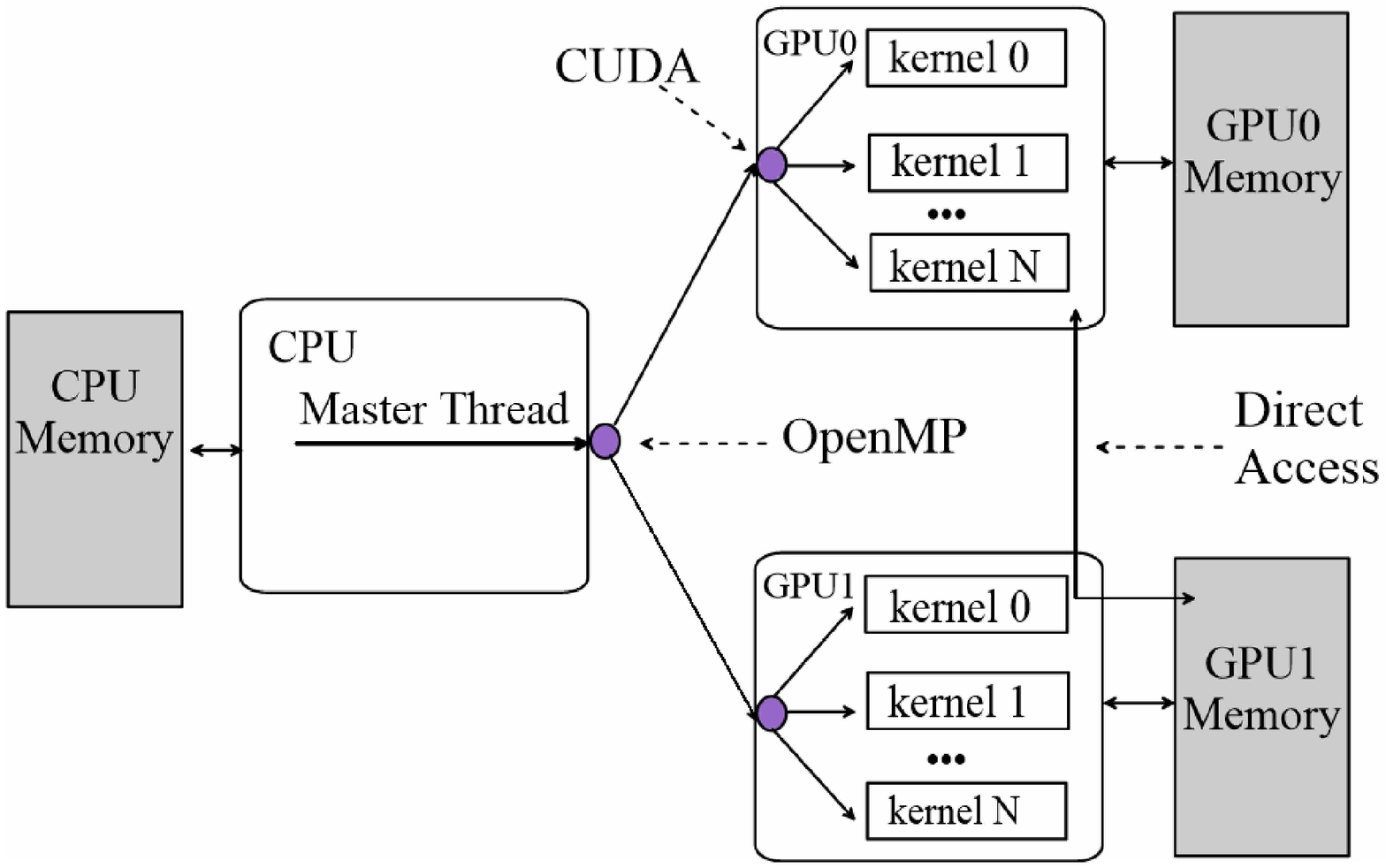}
 \vskip -0.321\hsize \hskip 0.6\hsize
            \includegraphics[width=0.4\hsize]{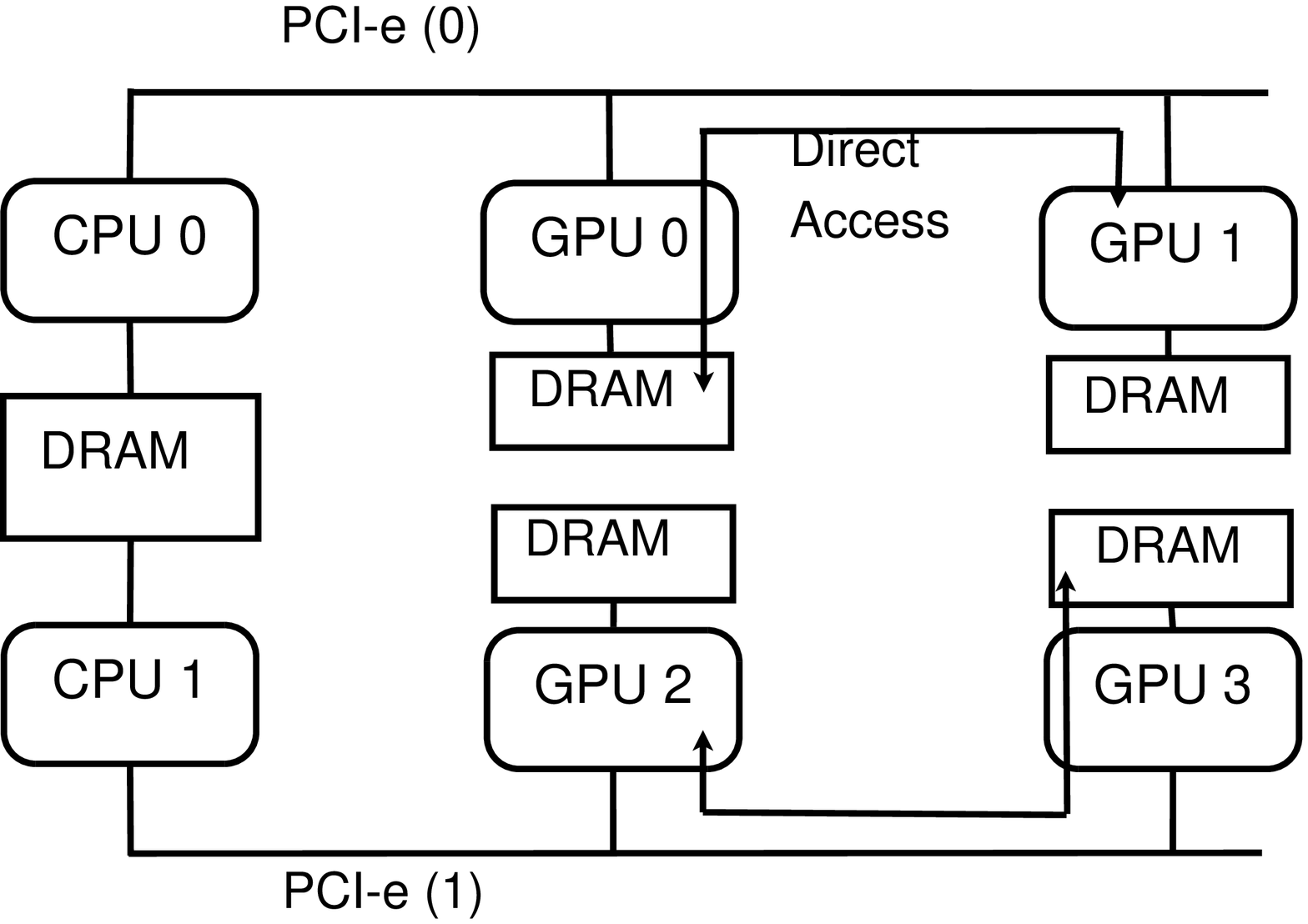}
\vskip  0.1\hsize \hskip 0.0\hsize
  \vbox{\hsize=0.999\hsize
  \caption{
a) The two-level scheme of parallelization with OpenMP--CUDA.
b) Architecture 2$\times$CPU+4$\times$GPU.
 }\label{Fig-CPU-GPU} }\vskip 0.0\hsize
\end{figure}

 The two-level parallelization scheme (Fig.~\ref{Fig-CPU-GPU}\,a) is more suitable for hybrid systems type CPU + $n\times$\,GPU.
  Direct Access technology provides the fast data exchange at different GPU.
   This technology is applicable only for the GPUs, which are connected to the PCI Express buses under the control of one CPU (Fig.~\ref{Fig-CPU-GPU}\,b).

\section{Comparison of the Effectiveness for Different GPU}

\begin{figure}[!h]
 \vskip 0.\hsize \hskip 0.0\hsize
            \includegraphics[width=0.43\hsize]{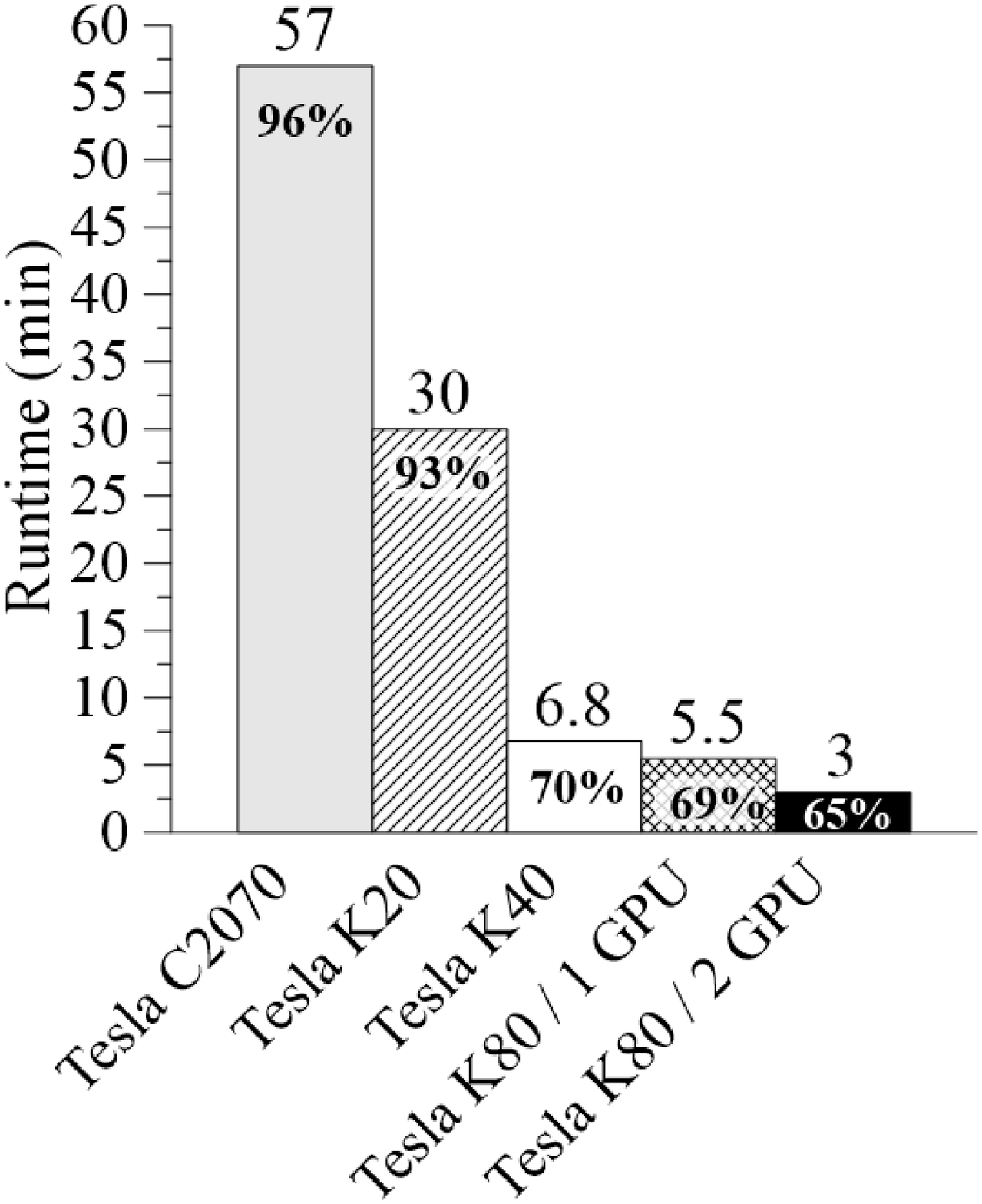}
 \vskip -0.445\hsize \hskip 0.510\hsize
            \includegraphics[width=0.49\hsize]{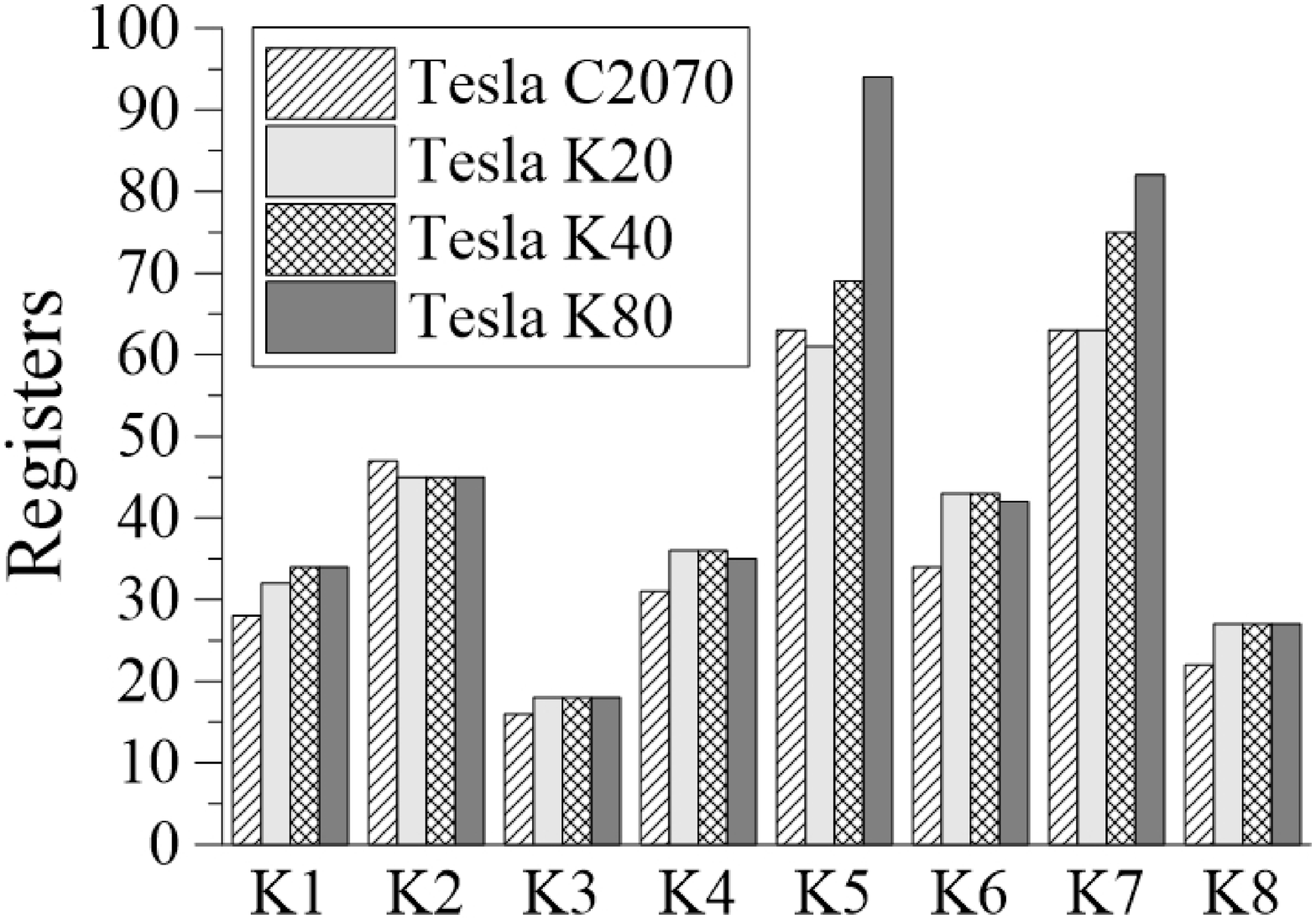}
\vskip  0.1\hsize \hskip 0.0\hsize
  \vbox{\hsize=0.999\hsize
  \caption{
a) Time calculation of flooding dynamics for the northern part of the floodplain for a period of 20 hours on different GPU using the grid 1024$\times$1024.
b) Distribution of memory registers of GPU-multiprocessors on CUDA-cores for different GPU.
 }\label{Fig-Runtime} }\vskip 0.0\hsize
\end{figure}

   The figure\,\ref{Fig-Runtime}\,a shows the diagram of Software Productivity for four Tesla graphics cards.
 The capture time executing CUDA kernels on the GPU (or GPU utilization) are quoted in percentages.
 We used the NVIDIA Parallel Nsight for profiling the program.
 In the transition to more efficient GPU with a large number of scalar cores we have a decrease in the percentage of GPU utilization and it is related to the number of computational cells $1024\times 1024$ (with $\Delta{x}=\Delta{y}=50$\,m).
GPU utilization increases in the case of $10-25$\,m spatial resolution.
 It is important to emphasize that the use of the spatial resolution $< 10$\,m may violate the approximation of the shallow water equations.
 Thus, the use of personal supercomputers with multi-GPU is the most suitable for the simulation of flooding over an area of about $10^4$\,km$^2$ considering the factor of GPU Utilization.

 The figure\,\ref{Fig-Runtime}\,b shows a diagram of the distribution of memory registers on a stream for CUDA-cores.
 We selected the parameters of the program for our graphic accelerators to avoid the spilling of registers.

\section{The Simulation Results}

\begin{figure}[!h]
\centering
\includegraphics[width=0.6\hsize]{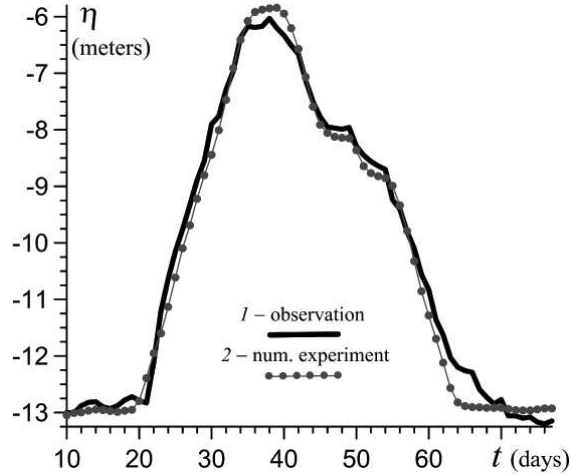}
\caption{ A comparison of water levels $\eta$ from observations (line \textit{1}) at gauging station ``Svetlyj Jar''
 with numerical simulation result (line \textit{2}).}
\label{Fig-obsnum}
\end{figure}

 We have a good agreement between the results of our numerical simulations and observation data for the dynamics of water level on the gauging stations and for the flooding area in May 2011 (figure\,\ref{Fig-obsnum}, See details in the \cite{khrapov2011,khrapov-pisarev2013,pisarev-khrapov2013}).
 As an example, consider the problem of emergency water discharge from the dam for the rate of $100000\,\textrm{m}^3/\,\textrm{sec}$.
Simulations were performed for the north part of the Volga-Akhtuba floodplain of the area $51200\,\textrm{m}\times 51200\,\textrm{m}$.
 Breaking wave formed  due to the emergency discharge leads to the complete flooding of the studied area of the floodplain for 20 hours.
The mean flow velocity in the floodplain is 5\,m/s and the average depth equals to 6\,m.
The ratio between the water and land in the entire field of modeling is $\sim 35$\%, and its maximum is $\sim 60$\%.
 For such problems, it is recommended to use the described approach (See paragraph 3) based on check of the presence of liquid in the computing CUDA-blocks, that speed up calculations by a factor of $1.5 - 2$.

\begin{figure}[!h]
 \vskip 0.\hsize \hskip 0.05\hsize
            \includegraphics[width=0.8\hsize]{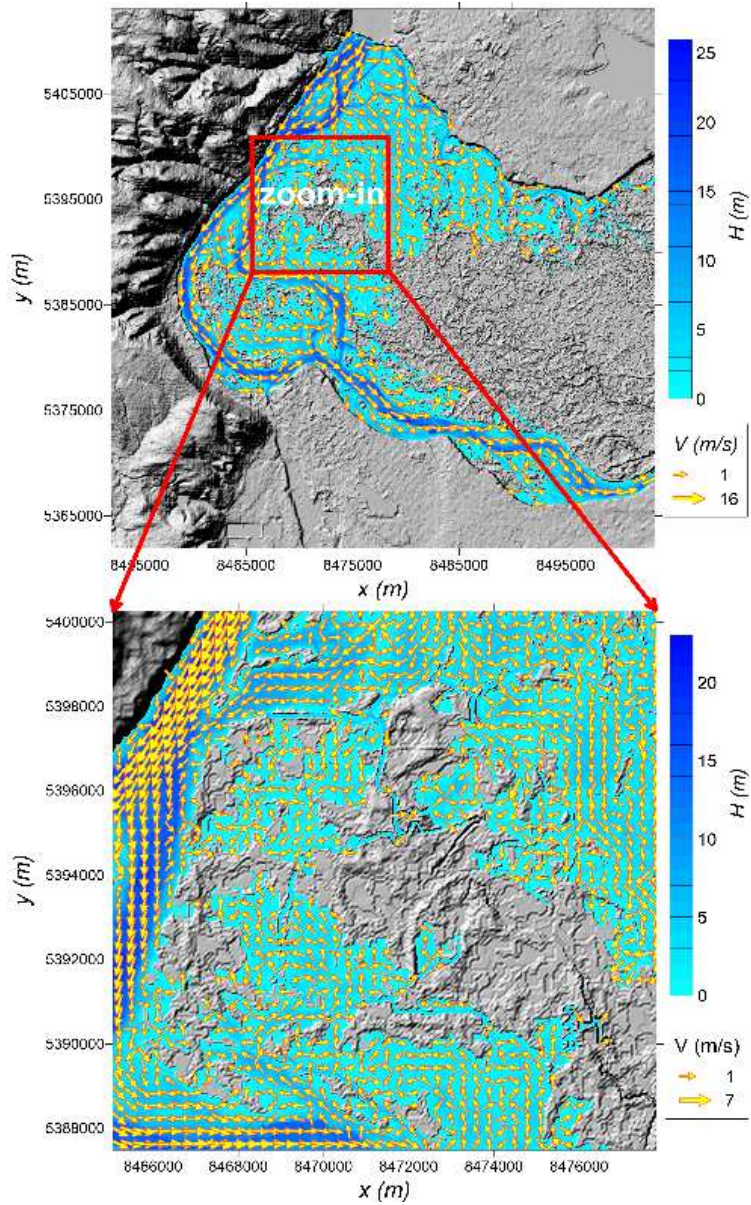}
\vskip  0.0\hsize \hskip 0.0\hsize
  \vbox{\hsize=0.999\hsize
  \caption {
  Hydrological state in the floodplain at time $t = 10$ hours using the zoom-in approach. Top frame shows the flow structure on the global grid. ($\Delta{x}=\Delta{y}=50$\,m), bottom frame shows the flow structure on the small grid ($\Delta{x}=\Delta{y}=12,5$\,m). The water depth distributions and the velocity field are shown  by the color and arrowheads respectively.
 }\label{Fig-ResultSimul} }\vskip 0.0\hsize
\end{figure}

The figure\,\ref{Fig-ResultSimul} shows the results of numerical modeling of flooding territories with taking into account of the zoom-in approach.
 Calculations were performed for the two grids:

 \noindent
 --- the global (main) grid which covers the entire area of simulations with the number of cells equal to $1024 \times 1024$ ($\Delta {x} = \Delta{y} = 50$\,m);

 \noindent
 --- the local grid for a critical region where we adopted a much higher spatial resolution $\Delta{x} = \Delta{y} = 12.5$\,m (size of calculation domain is $1024 \times 1024$, in the vicinity of village or complex terrain).

 \section{Conclusion}

We have investigated some features of the parallel implementation of numerical models for the Saint-Venant equations in the case when the flooded area changes in a wide range over the time. For example, the area under  the  water level may increase by a factor of tens or even  hundreds  in the period of spring floods for the Volga-Akhtuba floodplain. To improve the efficiency of these calculations we used a hybrid OpenMP-CUDA parallelization  approach and developed the method for choice of the CUDA-blocks with the aim to control the presence of liquid in the computational cells. Our parallel implementation reduced the computation time by a factor of $100-1200$ for different GPU in comparison to sequential code.

\subsubsection*{Acknowledgments.} The numerical simulations
have been performed at the Research Computing Center (Moscow
State University). AVK has been supported by the Russian Scientific Foundation (grant 15-02-06204), SSK is thankful to the RFBR (grant 16-07-01037). The simulation results part was developed under support from the RFBR and the Administration of Volgograd region grant 15-45-02655 (TAD). The study is carried out within the framework of government task by the Ministry of Education and Science of the Russian Federation (research work title No. 2.852.2017/PCH).

\end{document}